# Rigorous electromagnetic quasinormal-mode method made easy for users

Tong Wu[1,2,$] and Philippe Lalanne[1]*

[1]LP2N, CNRS, IOGS, Université Bordeaux, Talence, France

[2]Institute of Modern Optics, College of Electronic Information and Optical Engineering, Nankai University, Tianjin, China

[$]wutong1121@gmail.com

*philippe.lalanne@institutoptique.fr

**Abstract**: Full-wave numerical methods based on quasinormal modes (QNMs) offer valuable physical insights and computational efficiency for analyzing electromagnetic resonators. However, despite their advantages, many researchers in electromagnetism continue to favor real-frequency domain or time-domain approaches, often using finite element or finite-difference time-domain methods. This preference stems from various factors, including the perception that QNM theory is still developing or requires advanced mathematical tools from complex analysis. In this work, we combine numerical techniques with accurate approximations to simplify the computation of QNMs and enable ultrafast reconstructions using QNM expansions. The result is a new approach that is straightforwardly accessible to users familiar with real-frequency methods. We demonstrate the practicality of our approach through an open-source package [Doi: 10.5281/zenodo.18708748] implemented within a widely-used commercial photonics software.

## 1. Introduction

Micro- and nanoresonators play a central role in modern photonics, with their interaction with light fundamentally governed by the excitation of their natural resonance modes, commonly referred to as quasinormal modes (QNMs) in the broader literature on the complex analysis of non-Hermitian operators. When driven by an incident wave packet, these QNMs are excited, temporarily storing electromagnetic energy that is subsequently released.

Nearly 40 years ago, an innovative method for modeling electromagnetic resonator scattering was introduced by researchers at the Chinese University of Hong Kong [1,2]. This modal perspective provides a clear and intuitive understanding of resonator physics [3], as it enables the use of a reduced—sometimes extremely small—basis to study the temporal or frequency response. In other words, instead of solving the full problem across all degrees of freedom, the response can be accurately captured using a limited number of modes.

The Hong Kong group established important theorems on the orthogonality and completeness of QNM expansions for scalar fields in one-dimensional and spherical cavities in free space. However, a crucial aspect intrinsically tied to non-Hermiticity, namely, the QNM normalization, remained unresolved in this early work [4]. As a result, many analytical formulas derived within QNM theory, such as those for the modal excitation coefficients under an incident field or the complex frequency shifts of perturbed resonators, were approximately valid only in the limit of infinite quality factor, i.e., for nearly Hermitian systems.

The limitation of the original normalization was acknowledged by the Hong Kong researchers themselves (see Appendix A of [1]), but it was overlooked in subsequent works. Despite significant contributions addressing this issue [5-8], the error persisted within the electromagnetic QNM community [9], fostering skepticism among researchers outside the field regarding the validity of QNM-based approaches.

The normalization issue is now fully understood and resolved [4,10], so that the advancements made over the past decade have rendered the electromagnetic QNM formalism both mature and transparent [11]. It is therefore natural to see the emergence of electromagnetic QNM software packages [7,12-16]. To our knowledge, three groups are presently implementing such packages by leveraging either commercial or free Maxwell solvers using finite-element (FE) techniques. These solvers cover a broad range of geometries and materials commonly encountered in contemporary photonics. They have been quite widely adopted [17-21]. For example, the **MAN** (Modal Analysis of Nanoresonators) software [7,12,15], which was released in 2023 in a mature form, has been downloaded over 3,000 times across its various versions since 2013.

QNM-based methods are widely used in fields like mechanics [22-24], particularly in industries such as automotive engineering, where they are essential for efficiently modeling complex systems—even when hundreds of modes are needed for accuracy [25]. Despite their advantages, most researchers in electromagnetism continue to favor real-frequency domain or time-domain approaches, typically using FE or FDTD methods. In our view, real-frequency and time-domain methods are often less efficient and less physically transparent than QNM formulations, especially when only a few modes are sufficient for accurate results.

We believe this continued preference for direct approaches in electromagnetism is largely historical. Many engineers and researchers, especially in optics, receive limited training in complex analysis, and incorporating complex-frequency computations into FE or FDTD frameworks requires numerical expertise that may not always be readily available, particularly in experimental research groups. To facilitate broader adoption of QNM-based methods in the applied optics community, it could be highly beneficial to present these techniques in a more accessible and practical manner.

In addition to the mathematical challenges, there are several technical hurdles that must be overcome to fully leverage QNM-based analysis. First, although commercial FE software packages widely used in nanophotonics—such as CST Studio Suite and COMSOL Multiphysics—can compute electromagnetic QNMs, they often introduce significant errors for highly dispersive systems, particularly near electronic gaps or plasma frequencies. Second, expanding the optical response (e.g., near-field distributions or extinction spectra) in the

QNM basis requires extensive post-processing, such as computing the background incident field and evaluating frequency-dependent overlap integrals [3].

While these technical challenges can indeed be addressed [7,12-16], doing so typically demands advanced FE programming skills (e.g., modifying FE weak forms to ω-linearize the eigenvalue problem [12]) or expertise in software integration (e.g., coupling the FE solver with a programming environment). This complexity makes the workflow cumbersome, especially for non-experts.

To restore simplicity and transparency, we propose a two-step procedure. First, we demonstrate how a pole-search gradient descent algorithm [7]—capable of computing and normalizing QNMs—can be seamlessly implemented with Maxwell solvers operating in a standard platform with real frequencies. In the second step, we introduce several intuitive approximations to QNM theory to simplify the reconstruction of optical responses. These approximations not only introduce negligible errors but also significantly simplify and accelerate the reconstruction process in the QNM basis.

As a result, QNM analysis becomes straightforward and accessible to non-expert users familiar with real-frequency computations, without the need for advanced knowledge of complex analysis or finite element programming. We illustrate this approach using COMSOL Multiphysics and introduce a collection of standalone COMSOL models (MANlite), which implement the key components of QNM analysis. These include QNM computation and normalization, evaluation of complex mode volumes and quality factors [6,10,26], and calculating QNM excitation coefficients. All of these tasks are carried out directly within the FE environment, eliminating the need for external pre- or post-processing.

The remainder of this manuscript is structured as follows: Section 2 revisits the pole-search gradient descent algorithm, emphasizing key improvements that make the computation and normalization of QNMs more intuitive and user-friendly compared to previous approaches [7,15]. Section 3 presents representative examples demonstrating how normalized QNMs can be used to analyze the responses of electromagnetic resonators. This section introduces efficient approximations for quickly evaluating QNM excitation coefficients, a crucial quantity in coupled-mode theory [27] that describes how incident fields excite individual modes. We also show how these coefficients can be leveraged to reconstruct the extinction cross-sections of resonators in the QNM basis. Section 4 discusses the architecture and implementation of MANlite, an open-source software we developed to simplify the integration of our methods into FE simulations. Finally, Section 5 provides a conclusion to the manuscript.

## 2. Pole-search gradient descent algorithm

QNMs are source-free solutions to the Maxwell operator [3]

$$\nabla \times \tilde{\mathbf{E}}_m = -i\widetilde{\omega}_m \boldsymbol{\mu} \tilde{\mathbf{H}}_m, \tag{1}$$

$$\nabla \times \tilde{\mathbf{H}}_m = i\widetilde{\omega}_m \boldsymbol{\varepsilon} \tilde{\mathbf{E}}_m, \tag{2}$$

where $\tilde{\mathbf{E}}_m$ and $\tilde{\mathbf{H}}_m$ are the electric and magnetic QNM fields which are assumed to be normalized. $\boldsymbol{\varepsilon}$ and $\boldsymbol{\mu}$ are the spatially dependent permittivity and permeability tensors of the system, which comprises a resonator in a background medium with permittivity $\boldsymbol{\varepsilon}_b$. We assume hereafter nonmagnetic materials, $\boldsymbol{\mu} \equiv \mu_0 \mathbf{I}$, for simplicity. In Equations (1) and (2), $\widetilde{\omega}_m = \Omega_m + i\gamma_m$ is the complex-valued eigenfrequencies, $\Omega_m$ represents the resonant frequency, and $\gamma_m$ stands for the decay rate, encapsulating the lossy and leaky nature of the mode. We use the convention $\exp(i\omega t)$ throughout the paper.

The QNM fields of Equations (1) and (2) satisfy the outgoing wave conditions. We use perfectly matched layers (PMLs) which are traditionally used with real frequencies and can be as well used with complex frequencies [6,8]. PMLs are strictly valid when the background surrounding the resonator is invariant by translation, the most usual example being a layered substrate.

To ensure that the computed QNM fields are independent of the PML parameters, the thickness and damp parameters must be sufficiently large to guarantee that the outgoing wave fully decays to zero within the PML layer, without reflecting back into the physical domain. The complex frequencies result in QNM being divergent as $\exp(-i\widetilde{\omega}_m r/c)$ in the far field [11]. For low Q resonators (where $\text{Im}(\widetilde{\omega}_m) = \gamma_m$ is large), a larger PML thickness or higher damping parameter should be selected.

There are different methods to compute and normalize QNMs, as reviewed in Section 4 in [4]. Hereafter, we focus on the pole-search gradient descent method [7]. In Section 4, we will briefly consider another approach relying on conventional FE mode solvers. The strength of the pole-search gradient descent method lies in its generality: it can be applied to virtually any electromagnetic resonator, even those involving strongly dispersive materials or requiring satisfying the outgoing wave conditions in periodic media, such as the important case of resonators embedded in photonic-crystal waveguides [3]. It only requires that an analytical continuation of the permittivity and permeability through closed-form expressions (e.g., the Drude–Lorentz model or analytical functions fitted to measured real-frequency data) be known, which is a necessary step in complex analysis.

The pole-search gradient descent method computes and normalizes QNMs by locating poles in the complex frequency plane, starting from three initial guess values of the resonance frequency and iteratively refining the triplet toward the true resonance frequency. In its original version, the implementation of the iterative search in a FE software environment requires to pilot the software by external programs. For instance, with COMSOL Multiphysics, it requires MATLAB scripts [7,15], which many users find cumbersome.

We have successfully removed such a limitation by slightly modifying the algorithm. Specifically, we have introduced variables in the COMSOL model whose values are updated after each iteration. The external MATLAB is, therefore, not needed, since the definition, Maxwell equation solving, and post-processing are all performed in the model.

The new procedure is summarized in the flowchart of Figure 1. It relies on the fact that when a resonator is excited by a dipolar current source **J** oscillating at a complex frequency $\omega$ and located at $\mathbf{r}_0$; the optical response diverges as $\omega$ approaches to the resonance frequency (or pole) $\widetilde{\omega}$.

In the new procedure, one first begins by defining an initial triplet, $\{\omega_1, \omega_2, \omega_3\}_0$, consisting of three initial guess complex frequencies, $\omega_1, \omega_2, \omega_3$, chosen close to an expected QNM frequency. When the target QNM is well isolated from neighboring modes, the algorithm can successfully converge even from a relatively inaccurate initial guess. However, if QNMs are densely distributed within the frequency range of interest, a preliminary frequency sweep is required to obtain a more reliable estimate. This is important because, when attempting to identify all QNMs of interest by running the model multiple times, an inaccurate initial guess may cause some modes to be missed or lead to convergence to the same QNM in different runs. The sweep should therefore include a sufficient number of frequency points to clearly resolve spectral features—such as peaks or Fano resonances—associated with individual modes. An accurate initial guess is impossible with the frequency sweep if the response is dominated by a large number of modes whose spectral responses overlap and are difficult to identify. In such cases, the FE mode solver (see QNMEig discussed in Section 4), which is also integrated into MANLite, is recommended.

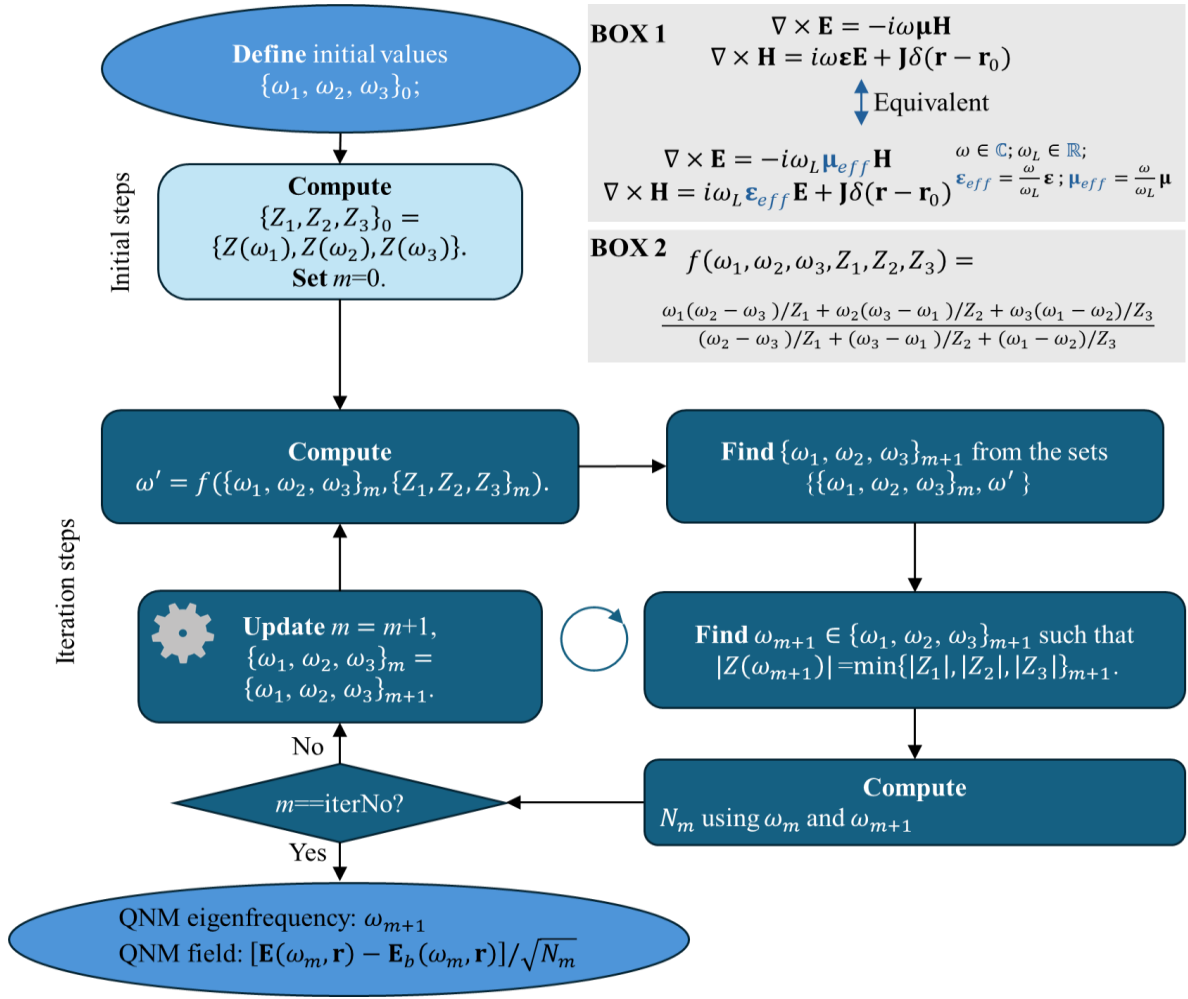

**Figure 1.** Flowchart of the pole-search gradient descent algorithm, implemented in COMSOL. State variables (marked with gears) are defined within the COMSOL model and updated iteratively based on the results of the previous calculation. **BOX 1:** The Maxwell equations at complex frequency are transformed into equivalent equations at real frequency by defining effective permittivity and permeability. **BOX 2:** Formula used for searching the pole.

In the initial stage, illustrated by the light-blue boxes in Figure 1, the program solves Maxwell equations for the three initial guess frequencies and computes the corresponding test fields: $Z_1 = Z(\omega_1)$, $Z_2 = Z(\omega_2)$, and $Z_3 = Z(\omega_3)$. The choice of the test function $Z$ is flexible: any optical response that takes the form of $Z(\omega) = a_0(\tilde{\omega}_m - \omega)$ for $\omega \approx \tilde{\omega}_m$ with $a_0$ a constant complex number can be employed. For example, we can set $Z(\omega) = 1/E_x(\omega, \mathbf{r}_t)$ where $E_x(\omega, \mathbf{r}_t)$ is the $x$-component of the complex-valued electric field at some test location $\mathbf{r}_t$.

The Maxwell equations are solved for a driving current source **J** located at $\mathbf{r}_0$ and oscillating at complex frequency are displayed in **BOX 1**. Since COMSOL restricts the computations to real-valued frequencies only, we introduce effective permittivity and permeability, $\boldsymbol{\varepsilon}_{eff} = \frac{\omega}{\omega_L}\boldsymbol{\varepsilon}$ and $\boldsymbol{\mu}_{eff} = \frac{\omega}{\omega_L}\boldsymbol{\mu}$, which recast the problem into equivalent Maxwell equations with identical electric and magnetic fields at a real frequency $\omega_L$ [7]. We must provide $\omega_L$, $\mathbf{J}$, $\mathbf{r}_0$, $\mathbf{r}_t$ and the maximum number *iterNo* of iterations. If convergence is achieved, the final result, i.e., the normalized QNM fields $[\widetilde{\mathbf{E}}_m, \widetilde{\mathbf{H}}_m]$ and the QNM eigenfrequency $\widetilde{\omega}_m$, is independent of these parameters.

After these initializations, the iterative process begins. This process involves repeatedly refining the initial frequency triplet $\{\omega_1, \omega_2, \omega_3\}_0$ and its associated test fields $\{Z_1, Z_2, Z_3\}_0$ in order to minimize the test field $Z$ for a specific complex frequency, which will ultimately approach the pole value.

The iterative steps are illustrated within the dark-blue boxes in Figure 1. We begin at step $m$, with the frequency triplet $\{\omega_1, \omega_2, \omega_3\}_m$, its corresponding test fields $\{Z_1, Z_2, Z_3\}_m$ and the expected pole value $\omega_m$, which corresponds to the frequency in the triplet that yields the minimum test value. A new complex frequency $\omega'$ is computed using the function $f$ described in **BOX 2**. The corresponding test field $Z(\omega')$ is also computed by solving Maxwell equations.

Next, among the triplet $\{\omega_1, \omega_2, \omega_3\}_m$, the algorithm selects the two frequencies with the smallest and second-smallest absolute values of the test field. By adding $\omega'$ to this pair, we form a new triplet $\{\omega_1, \omega_2, \omega_3\}_{m+1}$ at step $m+1$, along with its corresponding test fields $\{Z_1, Z_2, Z_3\}_{m+1}$. The new approximate pole value, $\omega_{m+1}$, corresponds to the frequency in the triplet that yields the minimum test value. The updates of the new triplets $\{\omega_1, \omega_2, \omega_3\}_{m+1}$ and $\{Z_1, Z_2, Z_3\}_{m+1}$ for the iteration time $m+1$ are shown with the gear highlighted box.

In most cases, $\omega_{m+1} = \omega'$. However, there are rare instances where $|Z(\omega')|$ is not smaller than the smallest absolute value of the triplet $\{Z_1, Z_2, Z_3\}_m$, in which case the curve $|Z(\omega_m)|$ may temporarily plateau. The updates of the new triplets $\{\omega_1, \omega_2, \omega_3\}_{m+1}$ and $\{Z_1, Z_2, Z_3\}_{m+1}$ for the iteration time $m + 1$ are shown with the gear highlighted box.

The iterative process goes on until the number of iterations $m$ reaches *iterNo*. In practice, convergence is typically achieved within 5 iterations, provided the frequencies of the initial triplet $\{\omega_1, \omega_2, \omega_3\}_0$ are already close to a pole. Although the number of iterations may increase if the initial guess is poor, in our experience, convergence is typically achieved within 10 iterations. If the guess values are too far away from the poles, the iterative process may not converge and is automatically stopped after a maximum iteration number assigned by the user.

Finally, it's worth mentioning that as $\omega'$ approaches to the QNM eigenfrequency, the Maxwell equation solver may fail due to the computed electric field reaching an extremely large value. To prevent this issue, we impose an upper bound value for the test field $1/Z(\omega')$ beyond which the updates of $\{\omega_1, \omega_2, \omega_3\}_m$ are halted.

**Numerical test.** We consider a typical geometry encountered in nanophotonics nowadays at optical frequencies [28,29], a silver nanocube antenna with side length 65 nm on a gold substrate coated with a thin polymer film (refractive index $n = 1.5$). Material dispersion is carefully taken into account. A single-pole Drude model ($\varepsilon(\omega) = \varepsilon_\infty - \varepsilon_\infty \frac{\omega_p^2}{\omega^2 - i\omega\gamma}$) with $\varepsilon_\infty = 1$, $\omega_p = 1.366 \times 10^{16}$ rad/s, and $\gamma = 0.0023\omega_p$ is used for the silver permittivity. The gold permittivity is defined by a double pole Drude–Lorentz model ($\varepsilon(\omega) = \varepsilon_\infty - \varepsilon_\infty \frac{\omega_{p1}^2}{\omega^2 - i\omega\gamma_1} - \varepsilon_\infty \frac{\omega_{p2}^2}{\omega^2 - i\omega\gamma_2 - \omega_{02}^2}$) with $\varepsilon_\infty = 6$, $\omega_{p1} = 5.37 \times 10^{15}$ rad/s, $\gamma_1 = 6.22 \times 10^{13}$ rad/s, $\omega_{01} = 0$, $\omega_{p2} = 2.26 \times 10^{15}$ rad/s, $\omega_{02} = 4.57 \times 10^{15}$ rad/s, and $\gamma_2 = 1.22 \times 10^{15}$ rad/s.

As shown in Figures 2a and 2b for the nanoparticle on mirror (NPoM) resonator, as $m$ increases, both the distribution of the normalized QNM field $|\tilde{E}_z|$ and the estimated pole frequency $\omega_m$ approach constant values, indicating convergence.

To compute physical quantities of interest, e.g. the complex mode volume shown in Figure 2c or the excitation coefficient (Section 3), the normalization factor must be evaluated. This factor, denoted as $N_m$ in the flow chart, is given by $N_m = \frac{-i\mathbf{J}\cdot[\mathbf{E}(\omega_m,\mathbf{r}_0)-\mathbf{E}_b(\omega_m,\mathbf{r}_0)]}{\omega_{m+1}-\omega_m}$. Its evaluation requires computing the electric field $\mathbf{E}_b$ radiated by the source $\mathbf{J}$ in the absence of the resonator (in the background layered substrate).

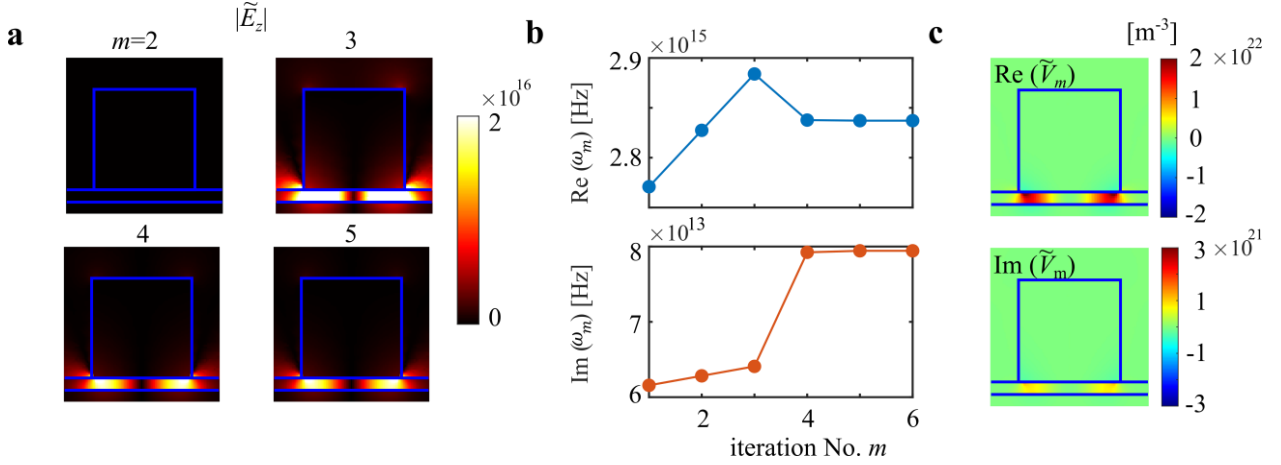

**Figure 2. Illustration of the pole-search gradient descent algorithm on the example of a NPoM resonator. a** Convergence of the normalized QNM field, $|\tilde{E}_z|$, during the iterations. **b** Convergence of Re($\omega_m$) and Im($\omega_m$). **c** Complex mode volume obtained after convergence.

## 3. Ultrafast reconstructions with QNM expansions

### 3.1 The QNM excitation coefficient

One of the central results of QNM theory is the QNM expansion of frequency-domain resonator responses. In this framework, the scattered field, $[\mathbf{E}_s(\mathbf{r},\omega),\mathbf{H}_s(\mathbf{r},\omega)]\exp(i\omega t)$, produced by a resonator under monochromatic excitation at frequency $\omega$, is expressed as a superposition of individual QNM contributions [1,3]

$$[\mathbf{E}_s(\mathbf{r},\omega),\mathbf{H}_s(\mathbf{r},\omega)] = \sum_m \alpha_m(\omega)[\tilde{\mathbf{E}}_m(\mathbf{r}),\tilde{\mathbf{H}}_m(\mathbf{r})], \qquad (3)$$

where the $\alpha_m(\omega)$'s are the modal excitation coefficients.

Equation (3) is valid for many material models, including Drude, multi-pole Drude-Lorentz ($\varepsilon(\omega) = \varepsilon_\infty - \varepsilon_\infty \sum_n \left[\frac{\omega_{p,n}^2}{\omega^2 - i\omega\gamma_n - \omega_{0,n}^2}\right]$), or N-pole Lorentz ($\varepsilon(\omega) = \varepsilon_\infty + \sum_n \left[\frac{A_n}{\omega-\omega_n} - \frac{A_n^*}{\omega+\omega_n^*}\right]$) permittivity models [30]. There is an abundance of possible excitation-coefficient expressions for $\alpha_m$ in literature. In a future publication, we will clarify why we chose [12]

$$\alpha_m(\omega) = \iiint_{\Omega_{res}} \left[\frac{\omega}{\widetilde{\omega}_m - \omega} \Delta\varepsilon(\mathbf{r}, \widetilde{\omega}_m) + \varepsilon_L(\mathbf{r}, \widetilde{\omega}_m)\right] \widetilde{\mathbf{E}}_m(\mathbf{r}) \cdot \mathbf{E}_b(\mathbf{r}, \omega) d^3\mathbf{r}, \qquad (4)$$

where $\Delta\varepsilon(\mathbf{r}, \omega) = \varepsilon(\mathbf{r}, \omega) - \varepsilon_b(\mathbf{r}, \omega)$ defines the permittivity contrast of the resonator within the resonator volume $\Omega_{res}$ (see Annex 2 in [3]) and $\mathbf{E}_b$ is the background electric field obtained by setting $\Delta\varepsilon = 0$. $\varepsilon_L$ is given by $\varepsilon_L(\mathbf{r}, \omega) = \varepsilon(\mathbf{r}, \omega) - \varepsilon_\infty(\mathbf{r})$. For non-dispersive materials, where $\varepsilon_L = 0$, Equation (4) remains valid.

For systems dominated by a few QNMs that can be strongly excited by incident light, Equation (3) can yield accurate results by computing fewer than 10 QNMs. However, in some cases, it may not achieve high accuracy with a small set of QNMs. For instance, near the material resonance frequency, the response is often significantly influenced by a large set of modes near accumulation points [12]. In weakly resonant systems, scattering can also be affected by a broad range of modes far from the spectral window of interest. Although technically possible [12], we do not recommend computing a large number of QNMs in these cases because the process is time-consuming. Rather, we suggest computing the few QNMs that capture the main resonant features and using interpolation for the smooth contribution of the non-resonant modes, which can be accurately computed with few number of real-frequency simulations [31].

**Approximation.** We now introduce a simplified approach in which the background electric field $\mathbf{E}_b$ in Equation (4) is evaluated at the real part of the QNM frequency, $\Omega_m = \text{Re}(\widetilde{\omega}_m)$. This leads to an approximate expression for the excitation coefficient $\alpha_m(\omega)$, given by

$$\alpha_m^R(\omega) = \left[\frac{\omega}{\widetilde{\omega}_m - \omega} \Delta\varepsilon(\widetilde{\omega}_m) + \varepsilon_L(\widetilde{\omega}_m)\right] \iiint_{\Omega_{res}} \widetilde{\mathbf{E}}_m(\mathbf{r}) \cdot \mathbf{E}_b(\mathbf{r}, \Omega_m) d^3\mathbf{r}. \qquad (5)$$

This substitution amounts to assuming that the spatial overlap between the incident and QNM fields within the resonator volume does not vary significantly as the frequency of the incident field changes over the spectral linewidth of the resonance. This assumption is generally well justified. Intuitively, this is because the excitation coefficients $\alpha_m(\omega)$ or $\alpha_m^R(\omega)$ take on large values only when $\omega \approx \text{Im}(\widetilde{\omega}_m)$, over a spectral interval on the order of $\text{Im}(\widetilde{\omega}_m)$. For plasmonic resonators, $\text{Im}(\widetilde{\omega}_m)$ can be relatively large; however, because the resonator is subwavelength, the background field $\mathbf{E}_b$ can be regarded as approximately constant over the resonator volume. Consequently, the overlap integral in Equation (4) is governed primarily by the amplitude of $\mathbf{E}_b$ at the origin of the coordinate. For plane-wave excitation, this amplitude exhibits no frequency dependence in a uniform medium and, in general, varies only slowly with frequency for illumination through a substrate. Conversely, for photonic crystal or ring resonators, the resonators are much larger, but the linewidth $\text{Im}(\widetilde{\omega}_m)$ is extremely small (quality factors often exceed 1000), so again $\mathbf{E}_b(\mathbf{r}, \Omega_m)$ and $\mathbf{E}_b(\mathbf{r}, \omega)$ remain very close. Conclusively, our approximation introduces only minor errors in general. A notable exception arises when the background field varies significantly across the QNM linewidth—for instance, in resonators placed on multilayer substrates whose optical thickness is comparable to the wavelength (see [32] for example).

To advance our goal of demonstrating QNM analysis directly from conventional Maxwell solvers, this approximation offers two key advantages. First, it greatly accelerates the computation of the excitation modal coefficients: as the frequency dependence is now analytically captured by the perfector in Equation (5), the spatial overlap integral needs to be computed only once, at the QNM frequency. Second, the method can be readily integrated into most FE Maxwell solvers. For instance, in COMSOL Multiphysics, it enables a fully self-contained model script that operates entirely within the software's post-processing environment, thereby eliminating the need for external MATLAB scripts (see Section 4).

**Numerical test.** To test the validity of the approximation, we consider the NPoM in Figure 2 illuminated by a TM-polarized plane wave with an incidence angle of 55°. The dominant QNMs in the visible spectral range are displayed in the complex frequency plane in Figure

3a. As we will see in the next subsection, typically three QNMs are enough to achieve an accurate reconstruction of the extinction cross section.

Here we introduce two key figures of merit [15] to identify the dominant QNMs: the Mode Ratio, $MR_m = \frac{\iiint_{\Omega_{phys}} \left|2\varepsilon(\widetilde{\omega}_m)+\widetilde{\omega}_m\frac{\partial \varepsilon}{\partial \omega}\right| |\widetilde{E}_m|^2 d^3\mathbf{r}}{\iiint_{\Omega} \left|2\varepsilon(\widetilde{\omega}_m)+\widetilde{\omega}_m\frac{\partial \varepsilon}{\partial \omega}\right| |\widetilde{E}_m|^2 d^3\mathbf{r}}$, and the Excitation Strength $ES_m = Q_m \iiint_{\Omega_{res}} |\varepsilon(\widetilde{\omega}_m)|^2 |\widetilde{E}_m|^2 d^3\mathbf{r}$, where $\Omega$ is the total computational domain, and $\Omega_{Phys}$ is the domain excluding the PML regions. The mode ratio, which ranges from 0 to 1, indicates whether an eigenstate is a true QNM or a PML-like mode. For true QNMs, the electromagnetic fields are concentrated within the physical domain with minimal penetration into the PMLs, causing the $MR_m$ to approach 1. Conversely, PML-like modes are characterized by a large field within the PMLs, resulting in an $MR_m$ typically less than 0.5.

The excitation strength quantifies how effectively a QNM can be excited by a plane wave or dipole source. It is proportional to the quality factor ($Q_m$) and the field intensity inside the resonator, which together determine the QNM excitation coefficient at its resonance frequency.

Figure 3a shows the position of the QNMs in the complex frequency plane, along with their quality factors $Q_m$ and frequencies $\widetilde{\omega}_m$. Their excitation strengths are represented by the disk sizes, while their mode ratios are indicated by the disk colors.

Figure 3b compares the results obtained from Equation (5) and those from the exact formulation in Equation (4), evaluated with the reconstruction toolbox available in [15]. The magnitudes of the excitation coefficients for the two dominant QNMs agree closely between the two approaches, confirming the validity of the approximation in typical scenarios.

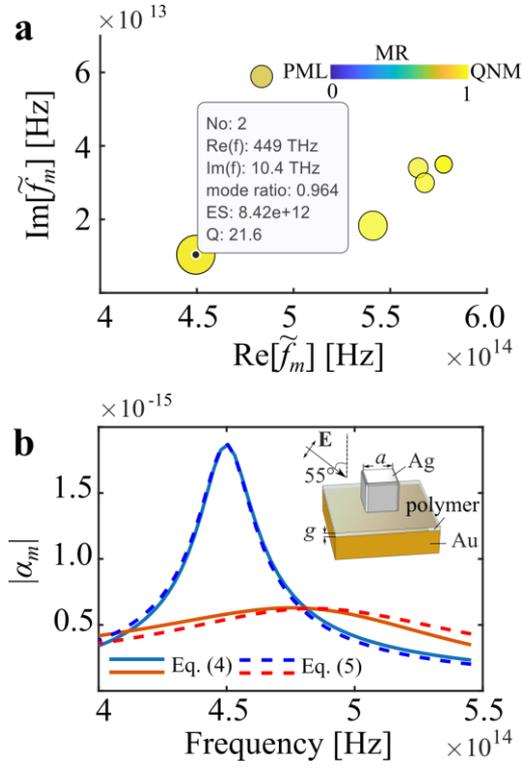

**Figure 3. QNM analysis can be directly performed using conventional real-frequency FE Maxwell solvers** for a representative classical photonic nanoresonator: a silver nanocube antenna, placed on an 8-nm-thick polymer film (refractive index 1.5) coated on a gold substrate. The cube has an edge length $a$ = 65 nm and is illuminated by a TM-polarized plane wave incident at 55°. **a** Complex frequency plane computed with MANlite. Important information on the QNMs that dominate the response in the visible spectral range are displayed: The size and color of the discs represent the excitation strength ($ES_m$) and mode ratio ($MR_m$). **b** Absolute values of the excitation coefficients $|\alpha_m|$ for the two dominant QNMs, computed using both the exact QNM theory (Equation (4)) and the approximation introduced in Equation (5).

### 3.2 The QNM contribution to extinction

Once the excitation coefficient $\alpha_m$ is determined, the scattered field can be reconstructed, enabling the evaluation of key physical observables, e.g. the extinction cross-section or the Purcell factor, expressed as superpositions of QNM contributions. In the following, we focus on extinction and introduce a new method for its computation. This approach, which is not available in existing QNM software [12-16], allows for the direct and unambiguous determination of the Fano coefficients associated with each resonance [33]. It thus eliminates the need to fit extinction spectra, a process that can be particularly challenging when multiple resonances overlap.

The extinction cross-section $\sigma_{ext}$ can be computed as an integral overlap between the total field, $\mathbf{E}_{tot}(\mathbf{r},\omega) = \mathbf{E}_b(\mathbf{r},\omega) + \mathbf{E}_s(\mathbf{r},\omega)$, and the background field $\mathbf{E}_b(\mathbf{r},\omega)$. We have

$$\sigma_{ext} = -\frac{\omega}{2I_0} \iiint_{\Omega_{res}} \text{Im}\{\Delta\varepsilon(\omega)\mathbf{E}_{tot}(\mathbf{r},\omega) \cdot \mathbf{E}_b^*(\mathbf{r},\omega)\} d^3\mathbf{r}, \qquad (6)$$

with $I_0$ the time-averaged Poynting vector of the incident plane wave in free space. Using Equations (4)-(5) and $\varepsilon_L(\omega)\mathbf{E}_{tot}(\mathbf{r},\omega) = \sum_m \alpha_m(\omega)\varepsilon_L(\widetilde{\omega}_m)\widetilde{\mathbf{E}}_m(\mathbf{r})$ (See Table 3 and Section 5.4 in [15]), the extinction can be decomposed into a sum of modal contributions

$$\sigma_{ext}(\omega) = \left[\sum_m \left(\sigma_m(\omega) + \sigma_{nr,m}(\omega)\right)\right] + \sigma_{nr}(\omega), \qquad (7)$$

with $\sigma_m = -\frac{\omega^2}{2I_0} \text{Im}\left[\frac{1}{\widetilde{\omega}_m - \omega} \zeta_m(\mathbf{E}_b,\omega)\, \zeta_m(\mathbf{E}_b^*,\omega)\right]$, $\sigma_{nr,m} = -\frac{\omega}{2I_0} \text{Im}[\zeta_m(\mathbf{E}_b^*,\omega)\, \zeta_m^L(\mathbf{E}_b,\omega)]$ and $\sigma_{nr} = -\frac{\omega}{2I_0} \text{Im}\left\{\iiint_{\Omega_{res}} \Delta\varepsilon_\infty \mathbf{E}_b^*(\mathbf{r},\omega) \cdot \mathbf{E}_b(\mathbf{r},\omega)\, d^3\mathbf{r}\right\}$. These expressions include a single resonant contribution, $\sigma_m$, and two non-resonant contributions, $\sigma_{nr,m}$ which depends on QNMs, and $\sigma_{nr}$, which does not. For simplicity, we have introduced the notations

$$\zeta_m(\mathbf{x},\omega) = \Delta\varepsilon(\widetilde{\omega}_m) \iiint_{\Omega_{res}} \mathbf{x}(\mathbf{r},\omega) \cdot \widetilde{\mathbf{E}}_m(\mathbf{r})\, d^3\mathbf{r}, \text{ and} \qquad (8)$$

$$\zeta_m^L(\mathbf{x},\omega) = \varepsilon_L(\widetilde{\omega}_m) \iiint_{\Omega_{res}} \mathbf{x}(\mathbf{r},\omega) \cdot \widetilde{\mathbf{E}}_m(\mathbf{r})\, d^3\mathbf{r}. \qquad (9)$$

Importantly, since extinction only involves first power term of the electric field, Equation (7) does not include any cross-terms between QNMs.

**Approximations.** Thus far, all expressions have been rigorously derived. We now reintroduce the same approximation as in the previous subsection, replacing $\mathbf{E}_b(\mathbf{r},\omega)$ by $\mathbf{E}_b(\mathbf{r},\Omega_m)$, results in closed-form expressions for the dependence of $\sigma_m(\omega)$, $\sigma_{nr,m}(\omega)$, and $\sigma_{nr}(\omega)$ on $\omega$. Specifically, we have:

$$\sigma_m^{(a)}(\omega) = -\frac{\omega^2}{2I_0} \text{Im}\left[\frac{1}{\widetilde{\omega}_m - \omega} \zeta_m(\mathbf{E}_b,\Omega_m)\, \zeta_m(\mathbf{E}_b^*,\Omega_m)\right], \qquad (10)$$

$$\sigma_{nr,m}^{(a)}(\omega) = -\frac{\omega}{2I_0} \text{Im}[\zeta_m(\mathbf{E}_b^*,\Omega_m)\, \zeta_m^L(\mathbf{E}_b,\Omega_m)], \qquad (11)$$

$$\sigma_{nr}^{(a)}(\omega) = -\frac{\omega}{2I_0}\text{Im}\left\{\iiint_{\Omega_{res}} \Delta\varepsilon_\infty \mathbf{E}_b^*(\mathbf{r},\Omega_m)\cdot\mathbf{E}_b(\mathbf{r},\Omega_m)\,d^3\mathbf{r}\right\}, \qquad (12)$$

where the superscript (a) highlights that the expression is obtained with an approximation. Under our assumptions, the extinction cross-section becomes

$$\sigma_{ext}^{(a)}(\omega) = \sum_m\left[\sigma_m^{(a)}(\omega) + \sigma_{nr,m}^{(a)}(\omega)\right] + \sigma_{nr}^{(a)}(\omega). \qquad (13)$$

Owing to our approximation, the $\omega$-dependent modal contribution of every QNM to the extinction cross section is known with a closed-form expression. Its evaluation requires computing only two overlaps integrals ($\zeta_m$ and $\zeta_m^L$) between the background field at frequency $\Omega_m$ and the QNM field, which requires very little computation once the QNMs are known.

**Numerical test.** Figure 4a compares the extinction contributions of the two dominant QNMs for the NPoM resonator. Solid lines correspond to $\sigma_m + \sigma_{nr,m}$ provided by the exact expression in Equation (7) while dotted lines correspond to the simplified form $\sigma_m^{(a)} + \sigma_{nr,m}^{(a)}$ in Equation (13). Despite the approximations — namely, evaluating $\mathbf{E}_b(\mathbf{r},\omega)$ at the real part of the QNM frequency $\Omega_m$ — the simplified formulation agrees very well with the full calculation, confirming its reliability for typical resonator configurations.

Figure 4b further demonstrates that, by summing the contributions of three dominant QNMs—two QNMs are shown in Figure 4a and the third one is symmetry-degenerate with one of two—the extinction is accurately reconstructed and closely matches the results of the real-frequency reference data obtained with COMSOL. Remarkably, the simplified expression in Equation (13) achieves accuracy almost as good as that of the exact formulation in Equation (7). Note that, in the present example, $\Delta\varepsilon_\infty = 0$, leading to $\sigma_{nr}^{(a)}(\omega) = 0$.

The QNM reconstruction method exhibits significantly improved computation efficiency. Once the QNMs are computed, the extinction across the entire spectrum shown in Figure 4b can be evaluated in less than 1 min. In contrast, the corresponding real-frequency (Exact) simulations require approximately 30 min.

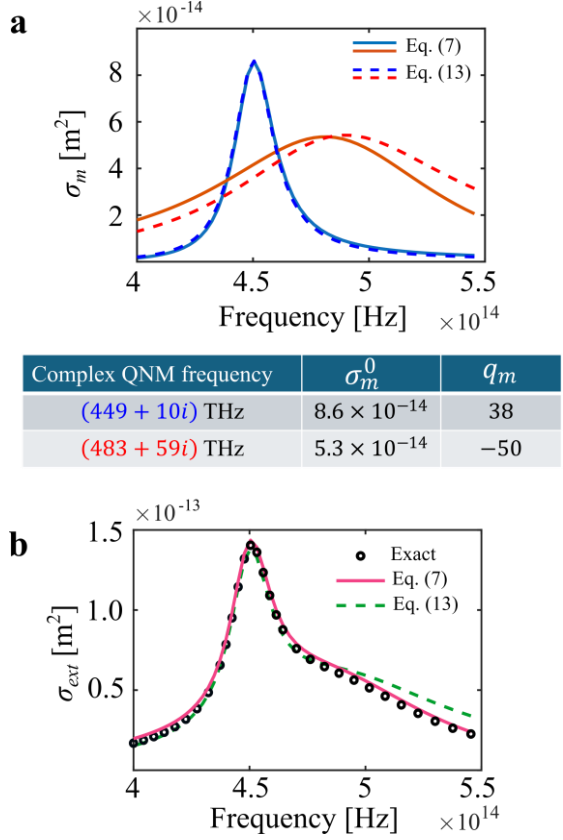

**Figure 4. a** Individual contributions of the two dominant QNMs to the extinction spectrum. The Fano factors $q_m$ and Fano intensities $\sigma_m^0$ for the two QNMs are listed in the table. **b** Comparison of extinction cross sections reconstructed using the three dominant QNMs with full-wave reference (exact) data obtained in the frequency domain with COMSOL. Reconstructions based on both Equations (7) and (13) show excellent agreement with the reference data shown as circles.

### 3.3 The Fano coefficient

In general, the Lorentz term $(\widetilde{\omega}_m - \omega)^{-1}$ has an absolute value much larger than 1 over the full linewidth of the resonance, with the response primarily dominated by the resonant term $\sigma_m^{(a)}$. Recently, it has been shown [33] that, by further approximating the factor $\frac{\omega^2}{2I_0}$ as $\frac{\Omega_m^2}{2I_0}$ in Equation (10), the resonant term $\sigma_m^{(a)}$ can be reformulated into the well-known Fano formula [34]

$$\sigma_m^{(a)-Fano} = \sigma_m^0 \left[ \frac{q_m^2 - 1 + 2q_m \Delta_m}{(\Delta_m^2 + 1)(q_m^2 + 1)} \right]. \tag{14}$$

This expression allows the cross section to be characterized by two real parameters, the Fano factor

$$q_m = \frac{\mathrm{Re}\{\xi_m^R\} + |\xi_m^R|}{\mathrm{Im}\{\xi_m^R\}}, \tag{15}$$

and the Fano intensity

$$\sigma_m^0 = \frac{\Omega_m^2}{2I_0} \frac{|\xi_m^R|}{\gamma_m}. \tag{16}$$

For convenience, we define $\xi_m^R = \zeta_m(\mathbf{E}_b, \Omega_m)\zeta_m(\mathbf{E}_b^*, \Omega_m)$ and $\Delta_m = \frac{\omega - \omega_m}{\gamma_m}$ with $\widetilde{\omega}_m = \Omega_m + i\gamma_m$ in Equations (14)-(16). Equations (15) and (16) are derived by substituting the factor $\frac{1}{\omega - \widetilde{\omega}_m}$ in Equation (10) with $\frac{1}{\omega - \widetilde{\omega}_m} = \frac{1}{\gamma_m}(\frac{\Delta_m + i}{\Delta_m^2 + 1})$, and comparing the reformulated equation with Equation (14).

The Fano factor $q_m \in [-\infty, \infty]$ provides a measure of the lineshape of the QNM contribution. When $|q_m| \to 0$, the profile is asymmetric (Fano resonance), while $|q_m| \to \infty$ corresponds to a symmetric Lorentzian lineshape. Importantly, the Fano intensity $\sigma_m^0 > 0$ quantifies the extinction strength. The values of the Fano factor and intensity for the QNMs of the NPoM resonator are displayed in the table of Figure 4a.

Traditionally, the Fano factor and intensity are obtained by fitting measured or simulated cross-section spectra. In contrast, Equations (15) and (16) enable their direct and accurate evaluation, eliminating the need to measure or compute the extinction spectrum.

## 4. MANlite implementation within COMSOL environment

To further assist users, alongside this report, we have released an open-source software package designed to demonstrate how to compute QNMs and utilize them to reconstruct scattered fields. The software, MANlite, includes eight COMSOL models that cover a broad spectrum of geometries, such as semiconductor photonic crystal cavities, 2D graphene nanodisks, plasmonic nanorods, multimode resonators, and NPoM resonators. In particular, the multimode resonator model demonstrates that the present method is not only valid for systems dominated by 1-2 QNMs but also capable of accurately reconstructing the response of multimode structures. Figure 5 shows a multimode resonator for which seven QNMs are required to accurately reconstruct the extinction. For highly multimode resonators, it is simpler to highlight only a few dominant contributions and to complement the reconstruction with this restricted by a few additional real-frequency simulations [31].

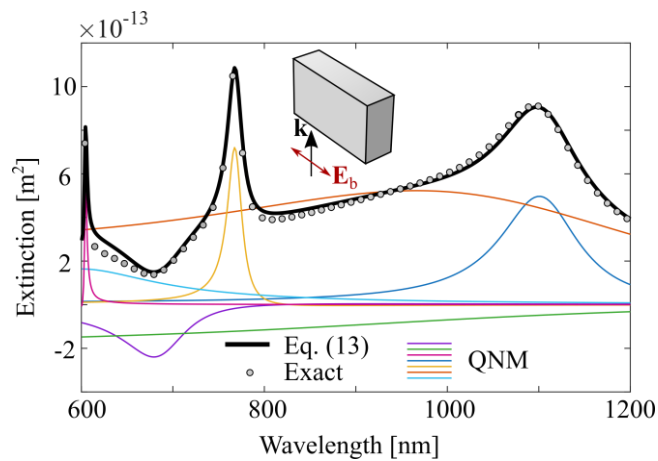

**Figure 5.** Comparison of extinction cross sections reconstructed using the 7 dominant QNMs with full-wave reference (exact) data obtained in the frequency domain with COMSOL. The multimode resonator is a $110 \times 220 \times 400$ nm$^3$ silicon box with a permittivity of 16, illuminated by a plane wave polarized along its long axis, as shown in the inset. The structure is inspired by Ref. [35].

Each model integrates the pole-search algorithm outlined in Section 2 and the ultrafast reconstruction methods detailed in Section 3. All models are fully self-contained and can be executed independently, without the need for any external MATLAB scripts.

The models are designed to be simple and transparent, enabling users with basic experience in real-frequency simulations using COMSOL to understand and modify them with ease. Additionally, users can build their own models by adapting the provided examples. In each model, variables are clearly categorized: those used for QNM computation and those used for ultrafast reconstruction. If the focus is solely on QNM computation, the reconstruction variables can be safely disregarded.

In these models, note that the triplets $\{\omega_1, \omega_2, \omega_3\}_m$ and $\{Z_1, Z_2, Z_3\}_m$ are not stored as standard variables. They are defined as COMSOL State variables, whose values are updated at every iteration.

For greater flexibility, MANlite also includes COMSOL models designed to perform QNM computations using the QNMEig solver [12], in addition to the pole-search algorithm described in Section 2. This solver takes advantage of the weak formulation—a general technique for analyzing mathematical equations that extends concepts from linear algebra to solve problems in fields such as partial differential equations. QNMEig relies on the eigenmode solver of COMSOL RF Module and provides enhanced performance for dispersive materials by incorporating auxiliary fields implemented through COMSOL Weak Form Module [15]. It operates entirely within the COMSOL environment, without the need for any external MATLAB scripts.

These models serve three key purposes. First, as validation tools, they enable users to cross-check results obtained with the pole-search algorithm. Second, they facilitate the computation of multiple QNMs without the need for an initial guess for the complex frequency. Finally, they function as educational resources, offering users a deeper understanding of how to formulate the weak form to effectively emulate finite-element methods by comparing the two solver implementations.

For readers interested in comparing the two solvers, Table 1 provides a summary of the key intrinsic quantities and their corresponding expressions in both model types. All MANlite models, whether using the pole-search algorithm or the weak form, are built with the COMSOL RF Module, which is why most variables are prefixed with emw. It is important to note that some variable notations may vary slightly depending on the solver used. However, when implemented correctly, both the pole-search and weak-form models yield identical results.

**Table 1.** QNM quantities that can be readily obtained with all COMSOL models in MANlite.

| Name | | Expression (QNMPole) | Expression (QNMEig) |
|---|---|---|---|
| $\widetilde{\omega}_m$ ($2\pi \widetilde{f}_m$) | complex QNM frequency | omega | QNM_omega |
| $N_m$ | Normalization factor | QN | |
| $Q_m$ | Quality factor | Re(omega)/(2Im(omega)) | emw.Qfactor |
| $\widetilde{E}_{m,z}$ | QNM **E** field, z-component | (emw.Ez-emwBg.Ez)/sqrt(QN) | emw.Ez/sqrt(QN) |
| $\widetilde{H}_{m,z}$ | QNM **H** field, z-component | (emw.Hz-emwBg.Hz)/sqrt(QN) | emw.Hz/sqrt(QN) |
| $\varepsilon(\widetilde{\omega}_m)$ | Permittivity at QNM eigenfrequency | epsilonA | |
| $\varepsilon_b(\widetilde{\omega}_m)$ | Background permittivity at QNM eigenfrequency | epsilonb | |
| $\mathrm{Im}(\widetilde{V}_m^{-1}) = \mathrm{Im}(2\varepsilon \widetilde{E}_{m,y}^2)$ | Imaginary part of the inverse of the mode volume, y-component | imag(2*(emw.Ey-emwBg.Ey)^2 /QN*epsilon0_const*epsilonA) | imag(2*(emw.Ey)^2 /QN*epsilon0_const*epsilonA) |

## 5. Conclusion

Over the past three decades, the popularity of quasinormal modes (QNMs) in electromagnetism has flourished [11,18-21,36–41], highlighting the remarkable diversity and vitality of the concepts. This sustained interest reflects not only the fundamental importance of resonances in modern photonics but also the growing recognition of modal analysis as a powerful framework for understanding and designing complex optical systems. We believe the field has now matured to a point where QNM analysis may become a standard tool in the design and optimization of optical resonators.

In this work, we aim to make QNM analysis more accessible to a wider community of researchers familiar with conventional real-frequency simulations. We address key numerical challenges faced when adopting QNM theory and introduce approximations that simplify optical response reconstruction while maintaining high accuracy. This results in a formulation that makes QNM computation and analysis both simple and intuitive. Our tests on contemporary nanophotonic examples suggest these approximations are generally valid across a wide range of important cases. Moreover, these approximations significantly speed up system response predictions, making QNM analysis particularly valuable during the initial design phases, where rapid exploration of parameter spaces is essential.

To verify the relevance of these developments, we have implemented an open-source software (MANlite) that encapsulates the proposed algorithms and approximations in a user-friendly environment. The software is fully integrated in COMSOL Multiphysics, a widely used platform in the photonics and microwave communities. While this integration may limit accessibility for researchers not familiar with COMSOL, the underlying methodology remains entirely transparent as all models can be viewed without a COMSOL license and can easily be adapted for use with other finite-element or frequency-domain solvers.

MANlite is available for download via Zenodo as part of the MAN distribution (for versions later than V9). We expect that this more user-friendly package will encourage broader adoption of QNM practices in the design and analysis of advanced nanophotonic devices.

**Acknowledgements**. The authors acknowledge helpful collaborations with Thomas Christopoulos, Marc Duruflé, Jianji Yang, Qiang Bai, and Mondher Besbes. Wei Yan at Westlake, Jean-Paul Hugonin and Christophe Sauvan at Institut d'Optique in Palaiseau, have all significantly contributed to establishing the QNM theory and implementing the software packages. PL acknowledges financial support from the UPiCO project (ANR-25-CE09-4019) and the European Union's Horizon research and innovation program under the Marie Skłodowska-Curie (QUARTE project, 101148330).

## 6. Disclosures

The authors declare no conflicts of interest.